\documentclass[prb,twocolumn,aps,superscriptaddress,showpacs,floatfix]{revtex4}
\usepackage{amsmath}
\usepackage{amssymb}
\usepackage{graphicx}
\usepackage{color}
\begin{document}

\title{Phase separation of antiferromagnetic ground states in systems with
imperfect nesting}

\author{A.L. Rakhmanov}
\affiliation{Advanced Science Institute, RIKEN, Wako-shi, Saitama,
351-0198, Japan}
\affiliation{Institute for Theoretical and Applied Electrodynamics, Russian
Academy of Sciences, 125412 Moscow, Russia}
\affiliation{Moscow Institute for Physics and Technology (State University), 141700 Moscow Region, Russia}

\author{A.V. Rozhkov}
\affiliation{Advanced Science Institute, RIKEN, Wako-shi, Saitama,
351-0198, Japan}
\affiliation{Institute for Theoretical and Applied Electrodynamics, Russian
Academy of Sciences, 125412 Moscow, Russia}

\author{A.O. Sboychakov}
\affiliation{Advanced Science Institute, RIKEN, Wako-shi, Saitama,
351-0198, Japan}
\affiliation{Institute for Theoretical and Applied Electrodynamics, Russian
Academy of Sciences, 125412 Moscow, Russia}

\author{Franco Nori}
\affiliation{Advanced Science Institute, RIKEN, Wako-shi, Saitama,
351-0198, Japan}
\affiliation{Department of Physics, University of Michigan, Ann
Arbor, MI 48109-1040, USA}

\begin{abstract}
We analyze the phase diagram for a system of weakly-coupled electrons
having an electron- and a hole-band with imperfect nesting. Namely, both
bands have spherical Fermi surfaces, but their radii are slightly
different, with a mismatch proportional to the doping. Such a model is used
to describe: the antiferromagnetism of chromium and its alloys, pnictides,
AA-stacked graphene bilayers, as well as other systems. Here we show that
the uniform ground state of this model is unstable with respect to
electronic phase separation in a wide range of model parameters.
Physically, this instability occurs due to the competition between
commensurate and incommensurate antiferromagnetic states and could be of
importance for other models with imperfect nesting.
\end{abstract}

\maketitle

\section{Introduction}

Electron models having a band structure with imperfect nesting are employed
to analyze properties of several physical systems. For example, such models
are used to describe: the antiferromagnetism (AFM) in Cr and its
alloys~\cite{RiceMod,Review88},
superconducting iron pnictides and iron chalcogenides
\cite{singh_physC2009,vorontsov_prb2010,gorkov_teit_prb2010,dagotto2012},
AA-stacked graphene bilayers
\cite{our_aa_paper,our_aa_phasep_preprint},
and other systems
\cite{gorkov_mnatzakov}.

\begin{figure}[btp]
\centering
\leavevmode
\includegraphics[width=8.5 cm]{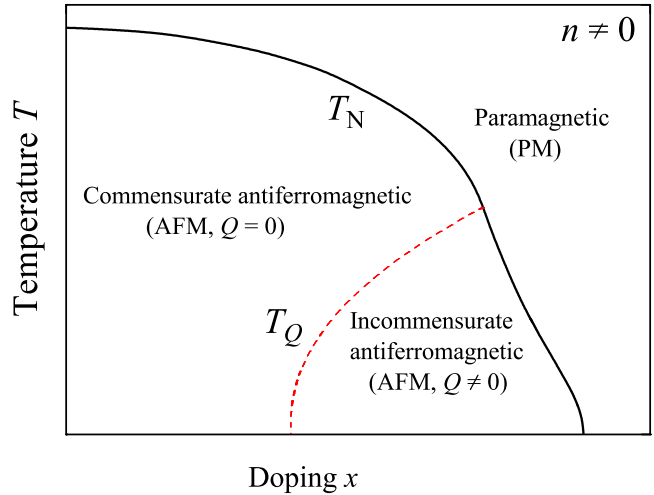}
\caption[]
{(Color online) A typical phase diagram of a magnetic system with
imperfect nesting in the 
$(x,T)$-plane, where $x$ is the doping and $T$ is the temperature. The
solid (black) curve represents the N\'{e}el temperature 
$T_\textrm{N}(x)$ 
separating the paramagnetic (PM) and the antiferromagnetic (AFM) states.
The dashed (red) curve is the boundary
$T_Q(x)$ 
between the commensurate 
($Q=0$) 
and incommensurate 
($Q\neq 0$) 
AFM states. This phase diagram is constructed for a finite density of
states of non-magnetic electrons ($n\neq0$, see the text).
}\label{gen_phase_diag}
\end{figure}

Postulating a spatially-homogeneous AFM ground state, a phase diagram for
models with imperfect nesting can be constructed. A typical 
$(x,T)$
phase diagram is schematically shown in
Fig.~\ref{gen_phase_diag}.
This diagram is split into three areas. The first is the paramagnetic
state at temperatures higher than the N\'{e}el temperature
$T_\textrm{N}(x)$.
If
$T<T_\textrm{N}(x)$
the system is in one of two magnetically-ordered states. The commensurate
AFM state exists at relatively low doping (where the nesting is good) and
higher temperatures, while the incommensurate AFM state appears at higher
doping and lower temperatures (see
Fig.~\ref{gen_phase_diag}).
Of course, in the commensurate phase the spatial variation of the order
parameter is commensurate with the crystal lattice period. As for the
incommensurate AFM, its structure is characterized by the wave vector
${\bf Q}(x,T)$.
This
${\bf Q}$
quantifies the smooth variation of the AFM order parameter in real space
over distances much longer than the lattice spacing. Since
${\bf Q}$
is zero in the commensurate phase, the condition
${\bf Q}(x,T)=0$
defines the boundary temperature $T_Q(x)$ between the commensurate and
incommensurate AFM.

The phase diagram in
Fig.~\ref{gen_phase_diag}
was obtained 
\cite{RiceMod}
assuming that the ground state of the system is uniform.
However, this assumption is not necessary valid: below we demonstrate that,
depending on the doping and the temperature, the homogeneous state may be
unstable. To prove this we calculate the chemical potential $\mu(x, T)$ for
the model Hamiltonian of the itinerant AFM proposed by
Rice~\cite{RiceMod}.
We observe that $\partial\mu/\partial x$ is negative in a considerable
portion of the $(x,T)$-plane. The homogeneous state compressibility,
therefore, is negative, and such state is unstable with respect to
electronic phase separation.

It is not difficult to check a particular model for the phase separation
instability: an interval of dopings where free energy is a concave function
of doping is the signature of phase separation. Yet, the phenomenon is
sometimes overlooked due to the fact that other important properties of a
homogeneous state bear little or no signature of the underlying
instability. For example, the single-particle gap of a homogeneous state
may be a smooth decreasing function of doping
\cite{our_aa_phasep_preprint},
and rises no suspicion that, in fact, the state is unstable. Thus, a
separate assessment of the thermodynamic stability is required.

In this study we will restrict ourselves to the Rice model. However, the
discussed mechanism for the phase separation could be of importance to
other systems with imperfect nesting. For example, there are experimental
indications that pnictides and chalcogenides may experience such a
phenomenon in some regions of doping and
temperature~\cite{Fe_separation_experiment}.

This paper is organized as follows.
In Sec.~\ref{homogeneous}
we introduce the model Hamiltonian and derive the equations describing the
homogeneous states of the model. The instability of the homogeneous states
are presented in
Sec.~\ref{instability}.
The discussion and conclusions can be found in
Sec.~\ref{discussion}.

\section{Main equations of the model}
\label{homogeneous}

We study the model proposed by
Rice~\cite{RiceMod}
to describe the incommensurate antiferromagnetism in chromium (see also the
review in
Ref.~\onlinecite{Review88}).
We mostly adhere to the notation of
Ref.~\onlinecite{RiceMod}.
However, to comply with modern conventions, some changes will be
introduced. We also correct some misprints present in the latter reference.

The model band structure has one spherical electron pocket and one
spherical hole pocket with different radii (imperfect nesting), as well as
other band or bands, which do not participate in the magnetic ordering.
All interactions are ignored except the repulsion between the electrons
in the ordering pockets.

The system we study is three-dimensional. Its Hamiltonian has the form
\begin{equation}
\label{Hamiltonian}
\hat{H}\!
=
\!\!\sum_{\mathbf{k},\sigma \atop \alpha = a, b, c}
	\epsilon^\alpha(\mathbf{k})n_{\mathbf{k}\sigma}^\alpha\!+\!
\frac{V}{\cal V}
\!\!\sum_{\mathbf{k}\mathbf{k'} \mathbf{q} \atop \sigma\sigma'}
a^\dag_{\mathbf{k}+\mathbf{q}\sigma}
	a^{\vphantom{\dagger}}_{\mathbf{k}\sigma}
	b^\dag_{\mathbf{k'}-\mathbf{q}\sigma'}
	b^{\vphantom{\dagger}}_{\mathbf{k'}\sigma'},
\end{equation}
where $\alpha$ is equal to either $a$ (electron pocket), $b$ (hole
pocket), or $c$ (non-magnetic bands). Also, $a^\dag$ ($b^\dag$) are
creation operators for the electron in the $a$ ($b$) pocket,
$n^\alpha$
is the number operator, $V$ is the Coulomb interaction, and
${\cal V}$
is the volume. The non-magnetic non-interacting $c$ band has finite density
of states
$N_r$
at the Fermi energy. The energy spectra for the electron and hole pockets
measured relative to the Fermi energy are taken as ($\hbar=1$)
\begin{eqnarray}
\label{spectra_ele}
\epsilon^a(\mathbf{k}) \!&=\!&
v_F\left(k\!-\!k_{Fa}\right)=v_F\left(k\!-\!k_{F}\right)\!-\!\mu, \\
\label{spectra_hole}
\epsilon^b(\mathbf{k}\!+\!\mathbf{Q}_0)\! &=&\!
-v_F\left(k\!-\!k_{Fb}\right)\!=\!-v_F\left(k\!-\!k_{F}\right)\!-\!\mu,
\end{eqnarray}
where
$k_F=(k_{Fa}+k_{Fb})/2$,
$\mu=v_F(k_{Fa}-k_{Fb})/2$
is the chemical potential, and the wave vector
$\mathbf{Q}_0$
connects the centers of the electron and hole pockets in reciprocal space.
We confine ourselves to the weak-coupling regime: $VN_m\ll 1$, where
$N_m=k_F^2/2\pi^2v_F$.

We treat Hamiltonian
\eqref{Hamiltonian}
with a mean-field approach. This is admissible since mean-field
approximations give accurate results for weakly-interacting electrons. The
starting point of our derivation is the case of perfect nesting, which
corresponds to
$\mu = 0$.
Under this condition, the radii of the electron and hole pockets are
identical. If we translate the electron pocket by the vector
${\bf Q}_0$,
its Fermi sphere coincides perfectly with the Fermi sphere of the hole
pocket.

Mathematically, the Hamiltonian
Eq.~(\ref{Hamiltonian})
is equivalent to the BCS Hamiltonian. Indeed, if we perform the following
transformations
\begin{eqnarray}
\label{to_BCS}
b^{\vphantom{\dagger}}_k \rightarrow b^\dag_k,
\qquad
b^{\dagger}_k \rightarrow b^{\vphantom{\dagger}}_k,
\end{eqnarray}
the interaction constant $V$ and the hole pocket dispersion
Eq.~(\ref{spectra_hole})
change sign.
The Hamiltonian for $a$ and $b$ bands becomes identical to two copies of
the BCS Hamiltonian. This mapping is very useful since it allows to use
the familiar BCS mean-field approach to study the
Hamiltonian~(\ref{Hamiltonian}).

Performing standard BCS-like calculations, it is easy to show that for
$\mu = 0$
the system is unstable to the ordering with the AFM order parameter
$\Delta_0=\frac{V}{\cal V}\sum_{\mathbf{k}}\langle
a^\dag_{\mathbf{k},\sigma}
b^{\vphantom{\dagger}}_{\mathbf{k}+\mathbf{Q}_0,-\sigma}\rangle$.
In the weak-coupling limit
\begin{eqnarray}
\label{Delta0}
\Delta_0 = \epsilon_F \exp (-1/N_m V) \ll \epsilon_F,
\end{eqnarray}
where
$\epsilon_F = v_F k_F$ is the Fermi energy.

The order parameter
$\Delta_0$
couples electrons with unequal momentum. Consequently, in real space the
order parameter
$\Delta_0$
corresponds to the rotation of the magnetization axis with wave vector
${\bf Q}_0$.
Since usually the $a$ and $b$ pockets are located in the high-symmetry
points of the Brillouin zone, the vector
${\bf Q}_0$
is related to the underlying lattice structure. Thus, this order may be
called commensurate.

Now consider the case of non-zero $\mu$. In such a situation the electron
and the hole Fermi spheres have different radii, and do not coincide upon
translation. However, the distance between the translated spheres
remains small if $\mu$ is small.

It is likely that $\Delta_0$ remains metastable for small non-zero $\mu$.
Yet, one may try to optimize the energy further by treating the translation
vector
${\bf Q}_1 = {\bf Q}_0 + {\bf Q}$
as a variation parameter. The new order parameter has the form:
\begin{eqnarray}
\label{Rice_order}
\Delta=\frac{V}{\cal V}\sum_{\mathbf{k}}\langle
a^\dag_{\mathbf{k},\sigma}
b^{\vphantom{\dagger}}_{\mathbf{k}+\mathbf{Q}_1,-\sigma}\rangle.
\end{eqnarray}
Unlike
${\bf Q}_0$,
whose magnitude is of the order of the magnitude of the primitive
reciprocal lattice vectors, the vector
${\bf Q}$
is small:
\begin{eqnarray}
|{\bf Q}| \sim |\Delta|/v_F \ll |{\bf Q}_0|.
\end{eqnarray}
Thus, the order parameter $\Delta$ describes order with a slowly-rotating
AFM magnetization axis. The real-space wavelength of the axis rotation is
equal to
$2\pi/|{\bf Q}|$.
This value is unrelated to the underlying lattice. Therefore, it is natural
to call such order incommensurate.

If the transformation
Eq.~(\ref{to_BCS})
is performed on a system with non-zero $\mu$, the Hamiltonian of our
magnetic system becomes the
Fulde-Ferrel-Larkin-Ovchinnikov (FFLO) Hamiltonian
\cite{ff,lo}
of a superconductor in the finite Zeeman field
$\mu_B H = - \mu$.
Moreover, the magnetic order parameter
(\ref{Rice_order})
becomes a superconducting order parameter. Superconducting order of this
type was first studied by Fulde and Ferrel
\cite{ff},
while Larkin and Ovchinnikov
\cite{lo}
considered the order parameter
$(\Delta {\rm e}^{i{\bf Q}_1 {\bf r}}
+ \Delta^* {\rm e}^{-i{\bf Q}_1 {\bf r}})$.
The latter order parameter periodically passes through zero in real space.
Below we will follow
Rice~\cite{RiceMod},
and use the Fulde-Ferrel-type order parameter
Eq.~(\ref{Rice_order}).
The Larkin and Ovchinnikov version, recently applied by Gor'kov and
Teitel'baum~\cite{gorkov_teit_prb2010}
to study the coexistence of the AFM and superconductivity in pnictides,
will be discussed in
Sec.~\ref{discussion}.

Equilibrium parameters of the system can be derived by minimization of
the thermodynamic potential
\begin{equation}\label{Omega}
\Omega=-T\ln
\left[
	{\rm Tr}\, e^{-\left(\hat{H}-\mu \hat{N}\right)/T}
\right],
\end{equation}
where
$\hat{N}$
is the operator of the total particle number, and
$k_B = 1$. 
To evaluate $\Omega$ in the mean field approximation, we need the
eigenenergies of the mean-field Hamiltonian. These are
\begin{eqnarray}\label{spectrum}
\nonumber
E_{1,2}(\mathbf{k})
&=&
\frac{\epsilon^b(\mathbf{k}+\mathbf{Q}_1)+\epsilon^a(\mathbf{k})}{2}
\\
&\pm&
\sqrt{
	\Delta^2
	+
	\left[
		\frac{
			\epsilon^b(\mathbf{k}+\mathbf{Q}_1)
			-
			\epsilon^a(\mathbf{k})
		     }
		     {2}
	\right]^2
    }.
\end{eqnarray}
Then the grand potential
$\Omega = \Omega (\Delta, Q, \mu, T)$
equals to
\begin{eqnarray}
\label{grand_pot}
\frac{\Omega}{\cal V}&=&
\frac{2 \Delta^2}{V}
-2 T \sum_{s=1,2} \int \frac{d^3{\bf k}}{(2\pi)^3}\,
{\rm ln} \left(
		1 + e^{-E_s({\bf k}, {\bf Q}, \Delta)/T}
	\right)\nonumber\\
&&-2 T N_r 
\int
{\rm ln} \left(
		1 + e^{-\left(\epsilon-\mu\right)/T}
	\right)d \epsilon\,.
\end{eqnarray}
Here the first and the second terms are the contributions of the ordering
bands, while the third term corresponds to the non-magnetic bands. To carry
out the integration over
${\bf k}$,
we expand the band energies in powers of
$|{\bf Q}|$
and
$\delta k = |\mathbf{k}| - k_F$:
\begin{eqnarray}
\epsilon^b(\mathbf{k}+\mathbf{Q}_1) + \epsilon^a(\mathbf{k})
\approx
2\mu + 2 Q \eta,
\\
\epsilon^b(\mathbf{k}+\mathbf{Q}_1) - \epsilon^a(\mathbf{k})
\approx
2 v_F \delta k + 2 Q \eta,
\\
\label{def_Q}
Q=\frac{v_F |{\bf Q}|}{2},
\end{eqnarray}
where
$\eta$ is the cosine of the angle between
${\bf k}$
and
${\bf Q}$.

Performing the integration, one finds the following expression for the
difference of the grand potentials in the AFM state
($\Delta \ne 0$)
and in the paramagnetic state
($\Delta=0$)
\begin{eqnarray}\label{DOmega}
\delta\Omega &=&
	\Omega (\Delta, Q, \mu, T)
	-
	\Omega (0, Q, \mu, T)=
\\
\nonumber
&&\!
\frac{
	k_F^3 {\cal V}
     }
     {
	\pi^2\epsilon_F
     }
\!\left\{
		\Delta^2\!
		\left(
			\ln{\frac{\Delta}{\Delta_0}}\!
			-\!
			\frac{1}{2}
\right)\!
+\!
\frac{Q^2}{3}\!+\!\mu^2\!+\!\frac{\pi^2T^2}{3}+\phantom{\int\limits_0^\infty}\right.\\
\nonumber
&&\left. T\!\int\limits_0^\infty\!\! d\xi\! \int\limits_{-1}^1\!d\eta \ln\!\left[ f(Q\eta\!-\!\mu\!-\!\epsilon)f(\mu\!-\!Q\eta\!-\epsilon)\right] \right\},
\end{eqnarray}
where
$\epsilon=\sqrt{\Delta^2+\xi^2}$,
and
$f(\epsilon)=1/[1+\exp(\epsilon/T)]$
is the Fermi function.

The equilibrium values of the gap $\Delta$ and the magnitude of the
structural vector $Q$ are determined by minimizing
$\delta \Omega$. 
Thus, they are solutions of the equations
$\partial (\delta\Omega)/\partial \Delta =0$
and
$\partial (\delta\Omega)/\partial Q =0$.
Differentiating
Eq.~\eqref{DOmega}
we derive straightforwardly
\begin{equation}\label{Delta}
\ln\frac{\Delta_0}{\Delta} =\int\limits_0^\infty\! \frac{d\xi}{2\epsilon}\!\int\limits_{-1}^{1}\!\!d\eta\left[f(\epsilon+\mu-Q\eta)+f(\epsilon-\mu+Q\eta)\right],\\
\end{equation}
\begin{equation}\label{Q}
\frac{2Q}{3}
\!=\!-\!
\int\limits_0^\infty \!\!d\xi
\!\!\int\limits_{-1}^{1}\!\!\eta \, d\eta
\left[
	f(\epsilon\!-\!\mu\!+\!Q\eta)+f(\epsilon\!+\!\mu\!-\!Q\eta)
\right],
\end{equation}
where $\Delta_0$ is given by
Eq.~(\ref{Delta0}).
For fixed values of $T$ and $\mu$, this system must be solved for $\Delta$
and $Q$.

Once $\Delta$ and $Q$ are found, the total number of electrons per unit
volume
$n(\mu)$
can be calculated. The latter quality is the sum of the numbers of
magnetic
$n_m(\mu)$
and non-magnetic electrons
$n_r(\mu)$.
The doping $x$ is defined as the difference
\begin{eqnarray}
x=n(\mu)-n(0)=n_m(\mu)+n_r(\mu)-n_m(0)-n_r(0).
\end{eqnarray} 
Since
$\mu,T\ll \epsilon_F$,
we have for the non-magnetic part
\begin{eqnarray}
n_r(\mu)=n_\mu(0)+N_r\mu.
\end{eqnarray} 
The number of electrons in the magnetic bands is given by
\begin{equation}
\label{x}
n_m(\mu)=
\frac{2}{\cal V}
\sum_\mathbf{k}
\left[
		f(E_1 (\mathbf{k}))+f(E_2 (\mathbf{k}))
\right].
\end{equation}
Thus, the doping is equal to
\begin{equation}\label{mu}
\frac{x}{x_0}
=
\frac{n\mu}{\Delta_0}
+
\!\!\int\limits_0^\infty\!\!\frac{d\xi}{2\Delta_0}\!\!
\int\limits_{-1}^{1}\!\!
d\eta\left[f(\epsilon-\mu+Q\eta)-f(\epsilon+\mu-Q\eta)\right],
\end{equation}
where 
$x_0=4\Delta_0N_m$ 
and 
\begin{eqnarray}
\label{n_def}
n=\frac{N_r}{2N_m}.
\end{eqnarray} 
Note that the integrals in
Eqs.~\eqref{Delta}, \eqref{Q}, and \eqref{mu}
can be evaluated exactly at
$T=0$. Thus, at zero temperature the integrals are replaced by
transcendental functions. The corresponding equations were derived in
\cite{ff,RiceMod}
using somewhat different notation. Notice, however, that zero-temperature
expressions of
Ref.~\onlinecite{RiceMod}
contain several misprints. For example, Eq.~(6) of 
Ref.~\onlinecite{RiceMod}
has an incorrect minus sign between function $G$ [cf. Eq.~(8) of
Ref.~\onlinecite{ff}
shows the correct plus sign]. Further, in the definition of $r_\pm$ a
factor of (1/2) must be placed in front of $Q$, see our
Eq.~(\ref{def_Q}).

\section{Instability of the uniform ground state}\label{instability}

Now we are ready to construct the phase diagram of our model as a function
of temperature and doping. The three coupled
Eqs.~\eqref{Delta},~\eqref{Q}, and~\eqref{mu}
are solved numerically. These determine
$\Delta(x,T)$, $Q(x,T)$, and $\mu(x,T)$.
Using these results we can calculate the N\'{e}el temperature 
$T_\textrm{N}(x)$,
for different values of $n$, as the lowest $T$ where
$\Delta= 0$.
The transition temperature $T_Q$ between the commensurate and
incommensurate AFM corresponds to the highest doping at which
$Q = 0$.
As a result, we obtain the phase diagram of the type shown in
Fig.~\ref{gen_phase_diag}.

\begin{figure}[btp]
\centering
\leavevmode
\includegraphics[width=8.5 cm]{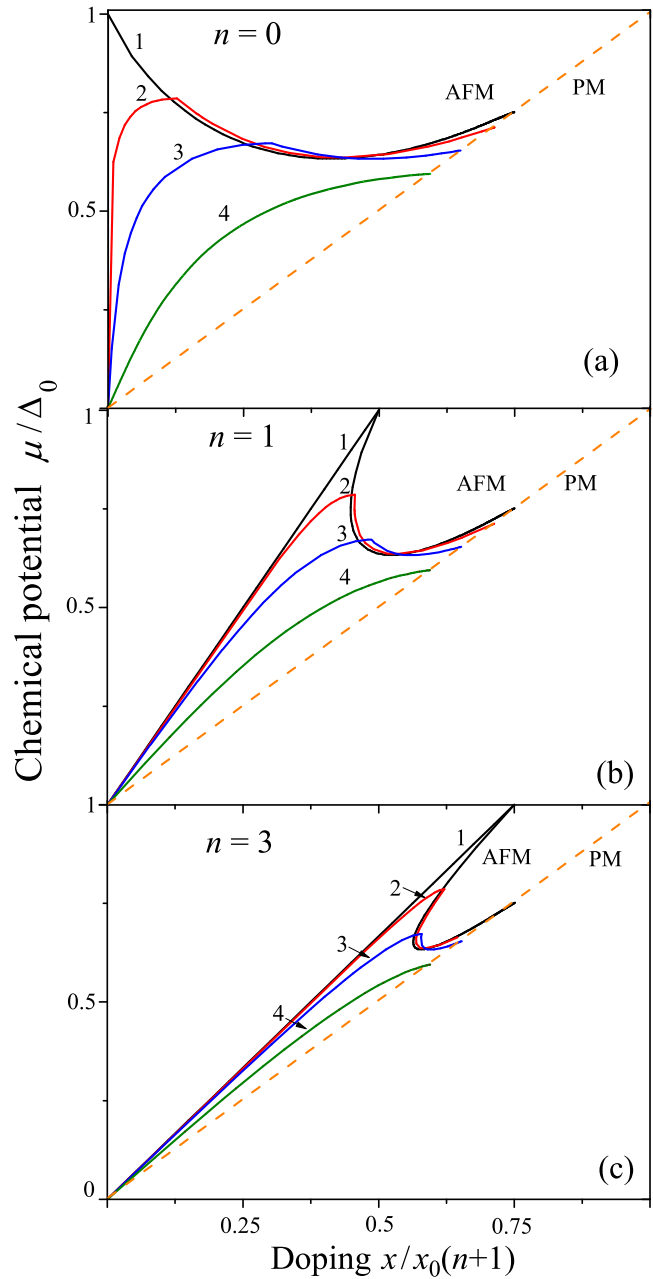}
\caption[]
{(Color online) Chemical potential $\mu$ versus doping $x$ for different
values of $n$ [see 
Eq.~(\ref{n_def})]: 
$n=0$
($a$), $n=1$ ($b$), and $n=3$ (c). The different temperatures considered
include: 
$T=0$ 
with solid (black) curves 1, 
$T/\Delta_0=0.1$ 
with solid (red) curve 2,
$T/\Delta_0=0.2$ 
with solid (blue) curve 3, and
$T/\Delta_0=0.35$ 
with solid (green) curve 4. Dashed (orange) lines correspond to 
$\mu(x)$ 
in the paramagnetic phase.
}\label{system}
\end{figure}

\begin{figure}[btp]
\centering
\leavevmode
\includegraphics[width=8.5 cm]{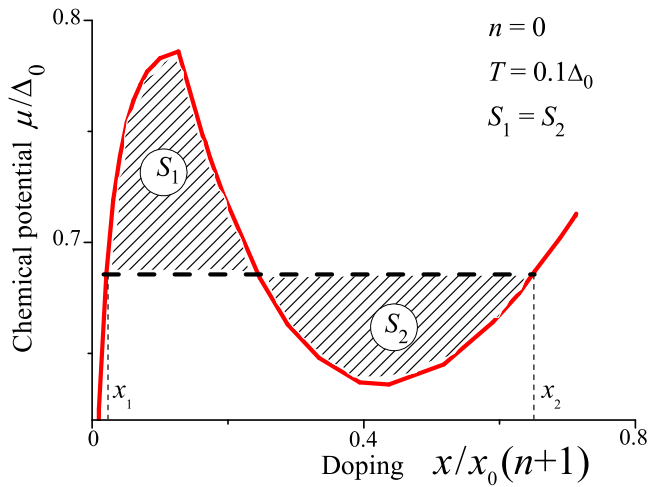}
\caption[]
{(Color online) Chemical potential $\mu$ versus doping $x$ for the
homogeneous phase, 
$T/\Delta_0=0.1$ 
and 
$n=0$ 
[solid (red) line]. The horizontal dashed (black) line shows the Maxwell
construction, with shaded areas
$S_1=S_2$.
}\label{system}
\end{figure}

However, constructing
Fig.~\ref{gen_phase_diag}
we assumed \textit{a priori} that the ground state of the model is uniform.
To check this assumption we plot the dependence of the chemical potential
on the doping, for different temperatures. For different values of $n$ the
results are shown in 
Figs.~2($a$, $b$, $c$).
Curves
$\mu(x)$
demonstrate three important features at temperatures lower than
$T^*\approx 0.317\Delta_0$, for any $n$.
First, the derivative
$\partial \mu/ \partial x$
is discontinuous at the transitions from commensurate to incommensurate
AFM
\cite{RiceMod},
and from incommensurate AFM to the paramagnetic phase. The second major
feature is that
$\mu(x)$
has three different values for a range of doping $x$ at low $T$ if
$n\gtrsim 1$,
which means that we have to choose the lowest energy 
solution~\cite{RiceMod}. 
Thirdly, the derivative
$\partial \mu/\partial x$
is negative at a finite range of doping.

This last peculiarity of
$\partial \mu/\partial x$
eluded the attention of previous studies; yet, it has very important
ramifications. Negative values of the derivative
$\partial \mu/\partial x$
mean that the compressibility,
$\kappa\propto \partial x/\partial \mu$,
is negative. This negative compressibility indicates that the homogeneous
state is unstable towards phase separation. 
In the phase-separated state
the system segregates into two phases with different doping values. Let us
denote these values as
$x_1$
and
$x_2$,
and the volume fractions of the corresponding phases as
$p_{1}$
and
$p_{2}$,
such that
$p_{1}+p_{2}=1$.
Then, the doping satisfies
$x=p_{1}x_1+(1-p_{1})x_2$, and $p_{1}=(x_2-x)/(x_2-x_1)$.

The values $x_1$ and $x_2$ can be found using the Maxwell
construction (see, e.g.,
Ref.~\onlinecite{LeBellac}).
Figure~3 illustrates the latter concept: the horizontal dashed line is
drawn in such a manner that the areas of the shaded regions,
$S_1$
and
$S_2$,
are equal:
$S_1=S_2$.

Using the Maxwell construction, the boundary
$T_{PS}(x)$
between the homogeneous and phase-separated states is calculated. This
boundary is shown by the dashed (red) curves in the
$(x,T)$
phase diagrams drawn in Fig.~4, for different values of $n$.

\begin{figure}[btp]
\centering
\leavevmode
\includegraphics[width=8.5 cm]{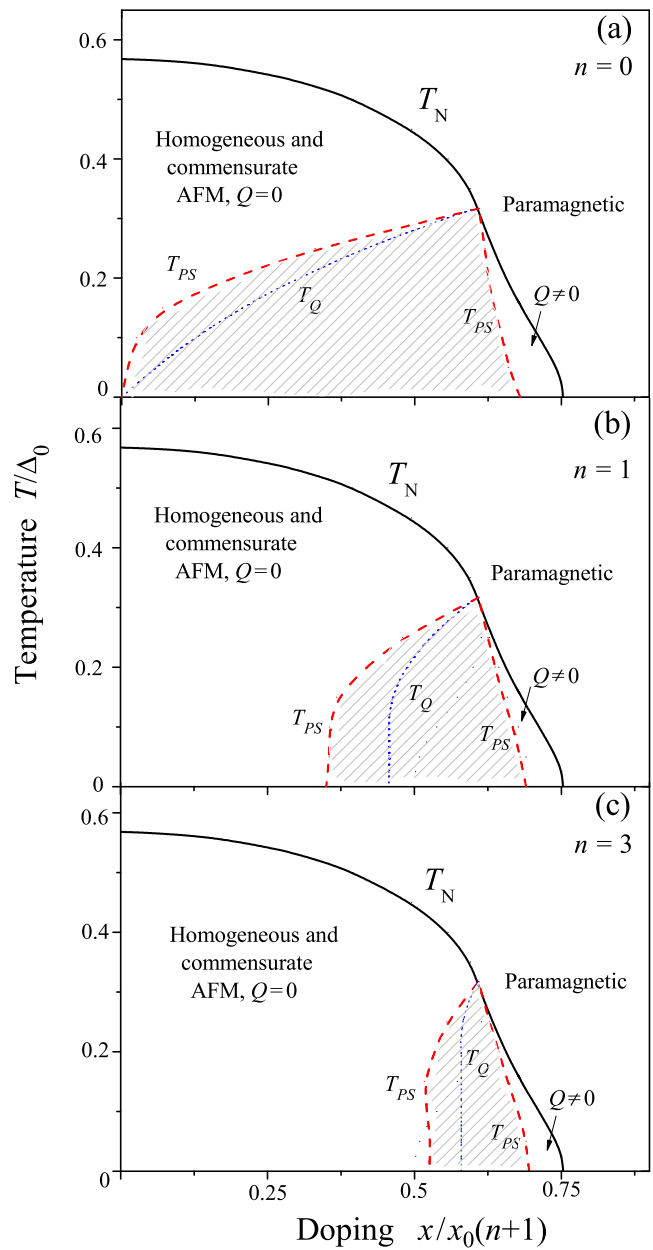}
\caption[] {(Color online) Phase diagram of the Rice 
model~\cite{RiceMod}
for 
$n=0$ ($a$),
$n=1$ ($b$), 
and 
$n=3$ ($c$).
The solid (black) curves represent the N\'eel temperature
$T_\textrm{N}(x)$, 
which separates the paramagnetic and AFM phases. The dotted (blue) curves
are
$T_Q(x)$, 
the boundary between the commensurate ($Q=0$) and incommensurate ($Q\neq 0$)
homogeneous AFM phases. The dashed (red) curve, $T_{PS}(x)$, is the
boundary between the uniform and phase-separated (shaded areas) phases.
}\label{system}
\end{figure}

The phase with lower doping, $x_1$, is the commensurate AFM ($Q=0$) while
the phase with higher doping, $x_2$, is the incommensurate AFM ($Q\neq 0$),
as it can be readily seen from Figs.~2,~3. Thus, here the phase separation
occurs due to the competition between two AFM states with different
structures. So, it is natural that the boundary temperature 
$T_Q$ 
lies between two lines 
$T_{PS}$ 
separating the homogeneous and inhomogeneous states (see Fig.~4). The phase
separation is absent for higher temperatures
$T>T^*\approx 0.317\Delta_0$.
This phase separation disappears simultaneously with the incommensurate AFM
phase. The area of the incommensurate AFM phase in Fig.~4 decreases, when
$n$ (which is proportional to the density of states in the non-magnetic
band) increases.  However, one must remember that in Figs.~2-4 the
horizontal scale changes when $n$ changes.

\section{Discussion}
\label{discussion}

In the previous section we demonstrated that the incommensurate AFM state
of the Rice 
model~\cite{RiceMod}
is unstable toward phase separation. This feature is likely to have
important consequences for the diverse set of materials where the nesting
degradation may be responsible for the destruction of the magnetic phase:
chromium and its alloys
\cite{Review88},
pnictides and chalcogenides
\cite{Fe_separation_experiment},
doped AA-stacked graphene bilayer
\cite{our_aa_phasep_preprint},
and others. Here we would like to discuss the obtained results and compare
these with other published works.

Above we used the mapping between the Rice model and the FFLO
superconductor. However, usually, the phase separation is absent from the
phase diagram of the FFLO superconductor. This has a very simple
explanation: these diagrams are plotted as a function of the temperature
and the Zeeman field (which is an analog of the chemical potential $\mu$ in
the Rice's model). The field, being an intensive thermodynamic quantity,
does not allow for phase separation. Instead, the system experiences a
first-order transition as a function of the Zeeman field
\cite{ff}.
However, for cold atoms in an optical trap it may be possible to control
not the field, but the polarization, which is an extensive quantity. In
this case, phase separation occurs
\cite{sheehy}.

As we mentioned above, besides
Eq.~(\ref{Rice_order}),
it is possible to consider other types of spatially-inhomogeneous order
parameters
\cite{stripes_luo,fflo_review}.
A particularly interesting possibility was discussed in
Ref.~\onlinecite{gorkov_teit_prb2010},
where the known numerical results for the two-dimensional FFLO
superconductor~\cite{bukh_ann_phys1994} 
with an order parameter of the Larkin-Ovchinnikov's type
was applied to study the Rice model. It was proposed
\cite{gorkov_teit_prb2010}
that, upon small doping, the order parameter in real space forms domain
walls. At a domain wall the gap vanishes locally, which makes these walls
a preferred place for the accumulation of the doped charge. It was
demonstrated 
\cite{bukh_ann_phys1994,gorkov_teit_prb2010}
that the formation of these domain walls becomes energetically
favorable for
\begin{eqnarray}
\mu > \mu_{DW} \approx 0.655 \Delta_0.
\end{eqnarray} 
The value of
$\mu_{DW}$
is somewhat smaller than
\begin{eqnarray}
\mu^{2D}_{PS} \approx 0.7 \Delta_0,
\end{eqnarray} 
which is the value of the chemical potential corresponding to the phase
separation in the two-dimensional Rice model at $T=0$. Since at low doping
the energy can be approximated by
\begin{eqnarray}
E = E_0 + \int_0^x\!\!\!\mu(x') dx' \approx E_0 + \mu(0) x,
\end{eqnarray}
where $E_0$ is the energy of the undoped state, we must conclude that at
low doping the phase-separated state is less favorable than the phase with
domain walls. However, the difference
\begin{eqnarray}
\mu_{PS}^{2D} - \mu_{DW} \approx 0.05 \Delta_0
\end{eqnarray} 
is small, and in real systems the balance may be shifted by the factors
unaccounted by the present model (e.g., anisotropy, Coulomb interaction,
disorder, etc.). Thus, the possibility of phase separation driven by the
mechanism proposed in this paper should be kept in mind when experimental
data are analyzed.

The experimental observation of phase separation in superconducting
pnictides and chalcogenides, which may be approximately described by the
Rice model
\cite{RiceMod}, 
was reported in several papers
\cite{Fe_separation_experiment}.
For example, Park and co-authors 
\cite{Fe_separation_experiment}
found the coexistence of magnetic and non-magnetic domains with a typical
size
$\sim 65$\, nm.
This observation is in general agreement with our proposed mechanism. While
our model predicts the separation into two magnetic phases, the AFM order in
the incommensurate phase is weak, and the energy is close to the energy of
the paramagnetic phase. Thus, either the AFM order parameter in the
incommensurate phase is below experimental sensitivity, or, being affected
by factors outside of our simple treatment, the weak phase itself is
replaced by the paramagnetic state. The latter scenario is quite likely,
given the small energy difference between the paramagnetic and
incommensurate AFM states. If the incommensurate AFM is destroyed, phase
separation occurs between the undoped commensurate AFM and the paramagnetic
states. Such type of phase separation was discussed for AA-stacked graphene
bilayers~\cite{our_aa_phasep_preprint},
whose band structure corresponds to the two-dimensional Rice model.

Let us briefly mention several complications not considered here, which,
nonetheless, may be present in experimental systems, and influence the
resulting phase diagram. In experiments, the electron and hole pockets may
have non-identical non-spherical shapes and different Fermi velocities. How
these factors affect the phase separation is a subject of further study.
In the weak-coupling regime,
Eq.~(\ref{Delta0}),
it is likely that sufficiently small deviations from the idealized model
will not affect qualitatively the outcome of the calculations, provided
that $V$ is not too weak.

Our calculations were performed in the weak-coupling limit. Would the phase
separation survive outside of this regime? Note that in order to discuss
the intermediate or strong-coupling regime, the Rice model is not a good
starting point. Rather, a multi-orbital lattice Hamiltonian is a better
approach. For such models the phase separation is a common 
phenomenon.~\cite{dagotto_book,other_pha_sep,PSmultiHubbard}
Further, recent numerical studies of the Hubbard model at intermediate and
strong 
coupling~\cite{trembley_hubb}
reported phase separation at finite doping. Thus, it is likely that
even for strong and intermediate interaction strengths it is possible to
find a region of the model's parameter space where phase separation occurs.
Although, in such regimes the mechanism of phase separation cannot be
described in terms of Fermi surface nesting.

To apply the theoretical results to experiments it is important to consider
the effects of disorder. We know that the FFLO state is very fragile with
respect to impurity scattering
\cite{takada}.
Thus, our incommensurate AFM, which is the mathematical analog of the FFLO
phase, is expected to be susceptible to microscopic imperfections.
Therefore, we conclude that the disorder-induced modifications to the phase
diagram is an open question which requires special attention.


The study of characteristic scales and geometry of the phase separated
state is beyond the scope of this study since it requires additional
information on the properties of the system, which are disregarded by the
Rice model. For example, even in the simplest approach to this problem, the
structure of the inhomogeneous state is governed by the interplay between
long-range Coulomb interaction and the energy of the boundary between
different 
phases.~\cite{Coulomb} 
Thus, it is reasonable to study the details of the phase-separated state
only if the particular physical system is specified in detail.

In conclusion, we have demonstrated that the uniform ground state of the
Rice model for an itinerant AFM with imperfect nesting is unstable with
respect to electronic phase separation in a significant range of dopings
and temperatures. In this range, the uniform system segregates into two AFM
phases, one of which is the commensurate AFM, while the second is the
incommensurate AFM. It is argued that such instability can occur in other
models with imperfect nesting because this effect is driven by the
competition between phases with different doping and different magnetic
structures.

\section*{Acknowledgements}

This work was supported in part by 
JSPS-RFBR Grant No.~12-02-92100, 
RFBR Grant No.~11-02-00708, 
ARO, 
Grant-in-Aid for Scientific Research~(S), 
MEXT Kakenhi on Quantum Cybernetics, and 
the JSPS via its FIRST program.
AOS acknowledges partial support from the Dynasty Foundation and 
RFBR grant No. 12-02-31400.

\end{document}